# Determining Viscosity of a Liquid with Smartphone Sensors: A Classroom-Friendly Approach Using Damped Oscillations

Sanjoy Kumar Pal[1,3], and Pradipta Panchadhyayee[2,3,*]

[1]Anandapur High School, Anandapur, Paschim Medinipur, India
[2]Department of Physics (UG & PG), Prabhat Kumar College, Contai, Purba Medinipur, India
[3]Institute of Astronomy Space and Earth Science, Kolkata -700054, India
[*]E-mail: pradipta@pkcollegecontai.ac.in

## Abstract

This study presents a classroom-friendly method for measuring the coefficient of viscosity of a liquid using a smartphone's accelerometer sensor. A metallic ball tied with a spring-mass system and submerged in mustard oil undergoes damped oscillations due to viscous forces. The Phyphox app is used to record the temporal variation of acceleration, from which the damping constant is calculated to determine the coefficient of viscosity of the oil. The experimentally obtained value is further validated using the Tracker app, and this value is shown to be in close agreement with the standard literature. This method provides an accurate, low-cost experiment ideal for educational settings, utilizing smartphone sensors for viscosity measurement.

## Theory and Experiment

The versatile applications of smartphone sensors are extensively used in physics experiments—that's why the dimensions of physics laboratories have radically changed in recent years. Advanced smartphone sensors like pressure and LiDAR sensors are used in day-to-day classrooms. A wide range of research articles has also been published on this subject [1-2]. Some classic smartphone sensors like accelerometer, sound, proximity, and magnetic sensors also make a valuable contribution to classroom teaching and education research [3-8].

In this article, we present a simple and innovative method to determine the viscosity of a liquid. A spring is suspended from a tripod stand, below which a small cotton bag containing a smartphone is tied with the spring. A metal ball with a radius 1.256 cm is hung from the bottom of the bag using a thread and made colinear with the axis of the spring. The metal ball is submerged in a cylindrical container (diameter = 11.1 cm and height = 22.2 cm) filled with mustard oil (see Fig 1a). By gently pulling the spring downward, the oscillations start with the vertical movement of the spring and the ball. When the ball oscillates within the mustard oil, special care has been taken to prevent it from touching the container's walls or bottom. In this regard, it should also be mentioned that the ball remains submerged in the oil and does not come into direct contact with the air during its motion. For small speeds of the ball, we can assume Stokes flow around the oscillating mass. So, the friction force experienced by the ball is $F = -6\pi\eta r v$, where $\eta$ is the viscosity of the oil, $r$ is the radius of the ball and $v$ is the velocity. The simple harmonic oscillations of the spring become damped [7-9] due to this viscous force. We note that the damping force acting on the ball is given by $F =$ $-bv$ (so in our Stokes flow, $b = 6\pi\eta r$). From the solution of the differential equation for a damped harmonic oscillator, the temporal dependence of the resulting displacement $x(t)$ can be written as:

$$x(t) = A \exp[-\gamma t] \sin(\omega t + \phi_0) \quad (1)$$

Where, $\gamma = \frac{2b}{m}$

This is valid for under-damped cases when the damping is small enough that the object can oscillate appreciably with time ($t$). In this equation $m$ is the total mass of the ball-bag-smartphone system, $\omega$ is the natural frequency of the spring oscillator, $\gamma$ is the damping constant, and $\phi_0$ is the initial phase.

Due to the viscous force, the amplitude of the oscillations decreases with time exponentially following the term, $\exp[-(\frac{2b}{m})t]$ . Therefore, if we measure the successive amplitudes of several oscillations with time, and plot a graph between $lnA$ vs $t$, we can determine the constant $(\frac{2b}{m})$. Using the equation $b = 6\pi\eta r.$ , we can calculate the coefficient of viscosity $\eta = \frac{m\gamma}{12\pi r}$.

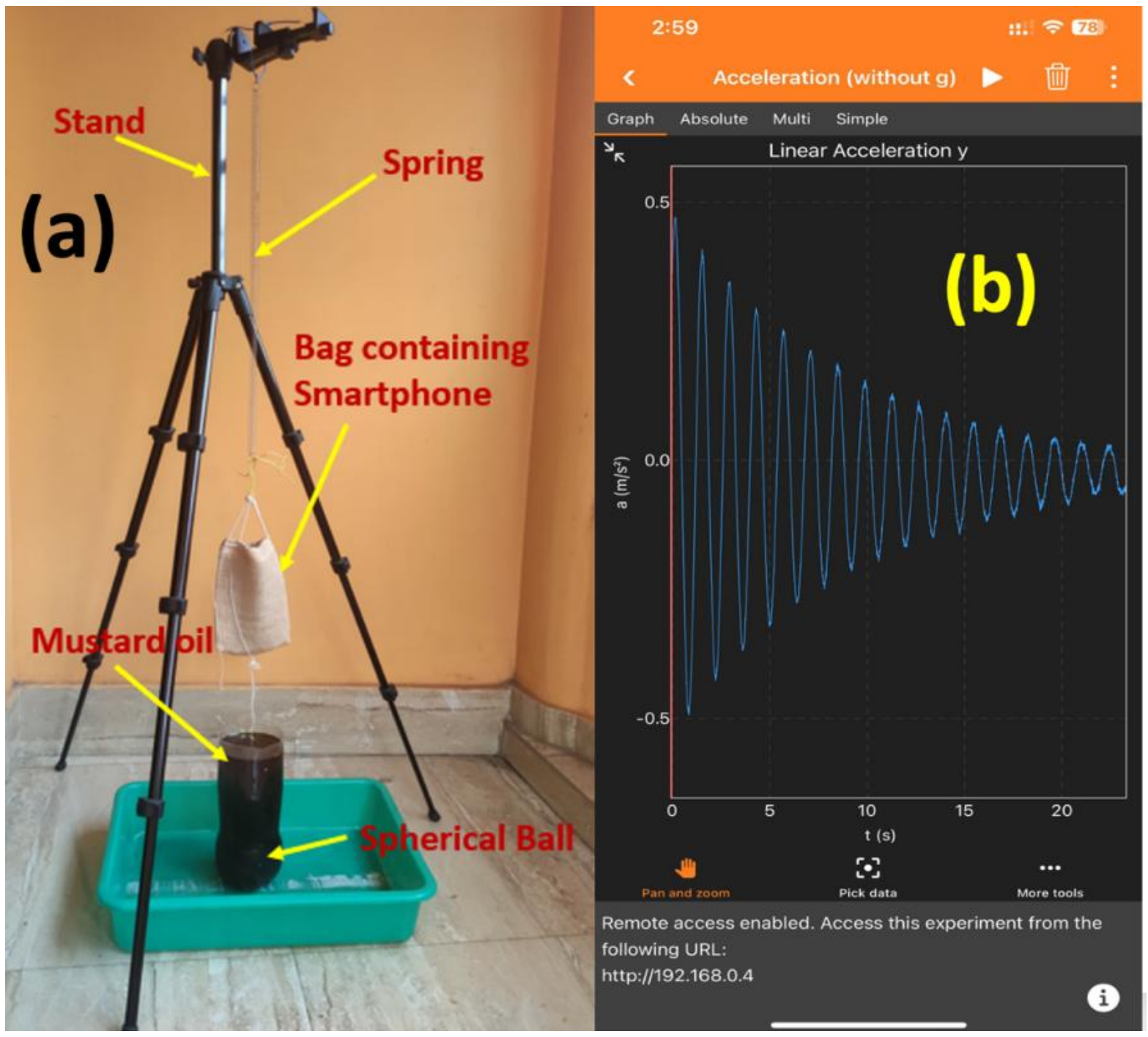

Fig 1: Set up of the experiment

In this experiment, we have used the accelerometer sensor via the Phyphox application on an iPhone 12 Pro Max to study the oscillatory motion of a spring-mass system. The Phyphox app is remotely operated from a MacBook to ensure precise data collection. The smartphone, positioned vertically in the small cotton bag, records the variation in acceleration along the $y$-axis. The accelerometer sensor captures data for acceleration with time in the coordinate axes ($x$-$y$-$z$). We have focused only on the $y$-axis to observe these variations (see Fig 1b). Although the sensor measures acceleration, we have measured the corresponding variations in the amplitude as a consequence of the direct proportionality between acceleration and amplitude. We record the peak coordinates of the acceleration curve over time, as provided by the Phyphox application (see Table 1), and subsequently plot the graph (see Fig. 2) of the logarithm of acceleration versus time (see Table 1). We have now computed the value of the damping constant $(2b/m)$ from this graphical presentation of the time dependence of acceleration. The negative slope of the resulting straight-line graph (Fig. 2) represents the damping constant, which is a key parameter in analyzing the damping behavior of the system.

Table 1: Time vs Acceleration data from Phyphox Accelerometer

| Time (s) | Acceleration ($a$) (m/s$^2$) | ln($a$) |
|---|---|---|
| 0.87 | 0.495 | -0.7032 |
| 2.24 | 0.427 | -0.8510 |
| 3.62 | 0.37 | -0.9943 |
| 5.02 | 0.322 | -1.1332 |
| 6.39 | 0.277 | -1.2837 |
| 7.81 | 0.251 | -1.3823 |
| 9.2 | 0.215 | -1.5371 |
| 10.6 | 0.197 | -1.6246 |
| 12 | 0.168 | -1.7838 |
| 13.4 | 0.151 | -1.8905 |
| 14.75 | 0.145 | -1.9310 |
| 16.12 | 0.12 | -2.1203 |

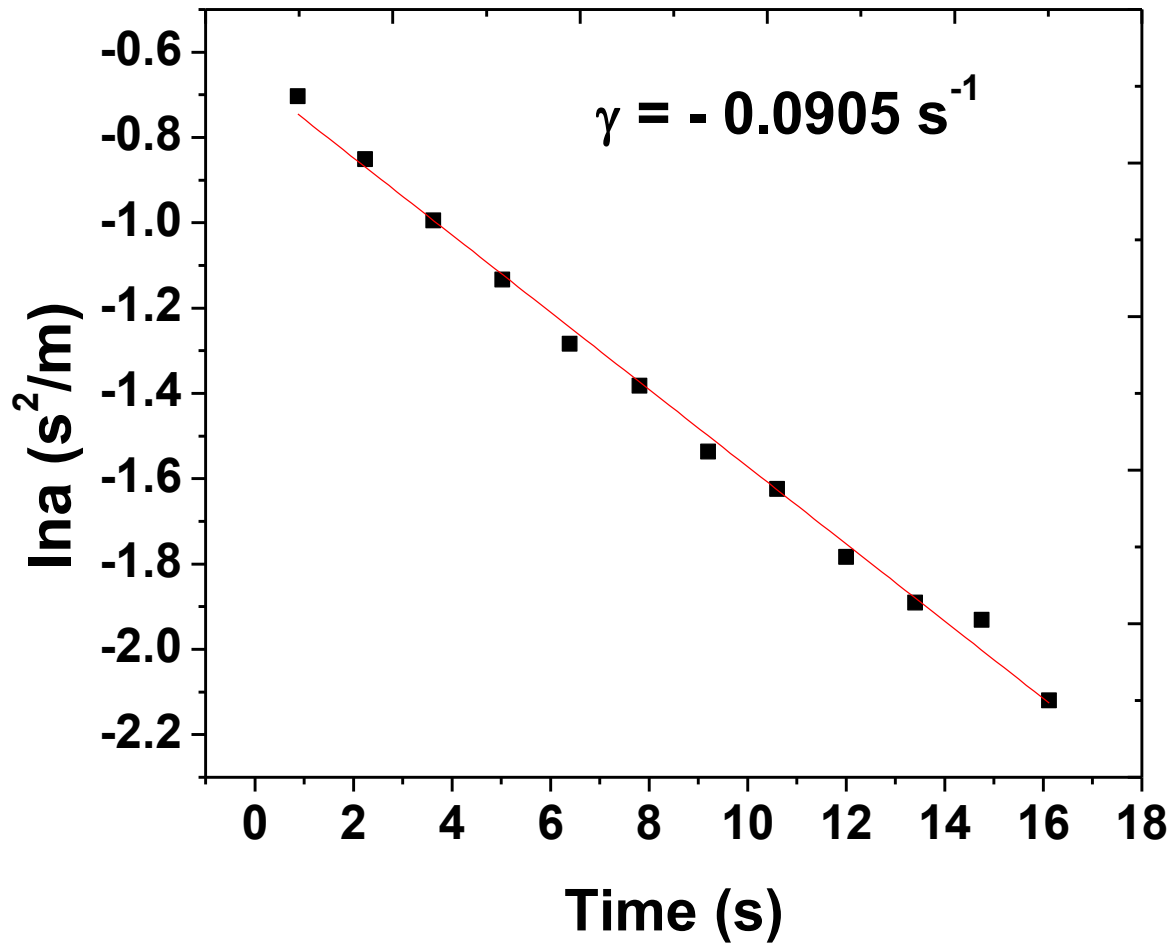

Fig 2: Graph between ln ($a$) vs time ($t$) for Phyphox method

From the Fig. 2, we observe that the slope of the straight-line graph is -0.0905 $s^{-1}$. The total mass of the spring-smartphone-ball system is measured to be 328.2 g. Using this information, we calculate the value of the damping coefficient $b$ to be 14.851 $gm.s^{-1}$. Consequently, the value of the coefficient of viscosity of the mustard oil is determined to be 0.627 poise, which is in close agreement with the literature value, 0.630 poise [10].

We have also verified the value obtained using the Tracker application [9]. In this setup, a spring is suspended from the support of the tripod stand and connected with a ball (m = 77.1 g) at the bottom using the same thread. Here, like in the previous case, the ball is kept fully submerged in mustard oil. The ball oscillates due to the motion of the spring and suffers damping in oscillations due to viscous friction. We have recorded the videos of oscillations using a smartphone and transferred those to a laptop for analysis with the Tracker application (see Fig 3).

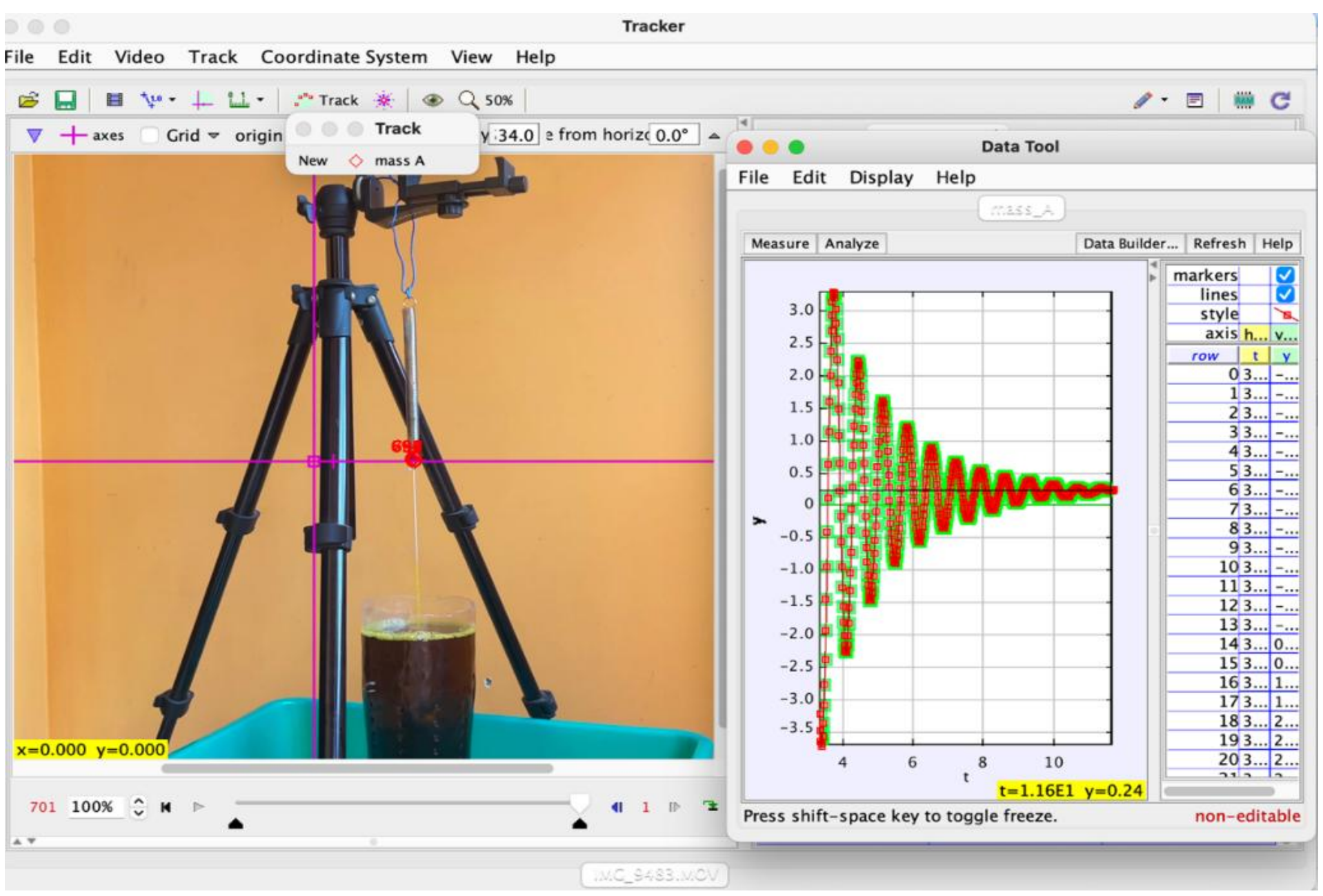

Fig 3: Screenshot at the time of analyzing by tracker application

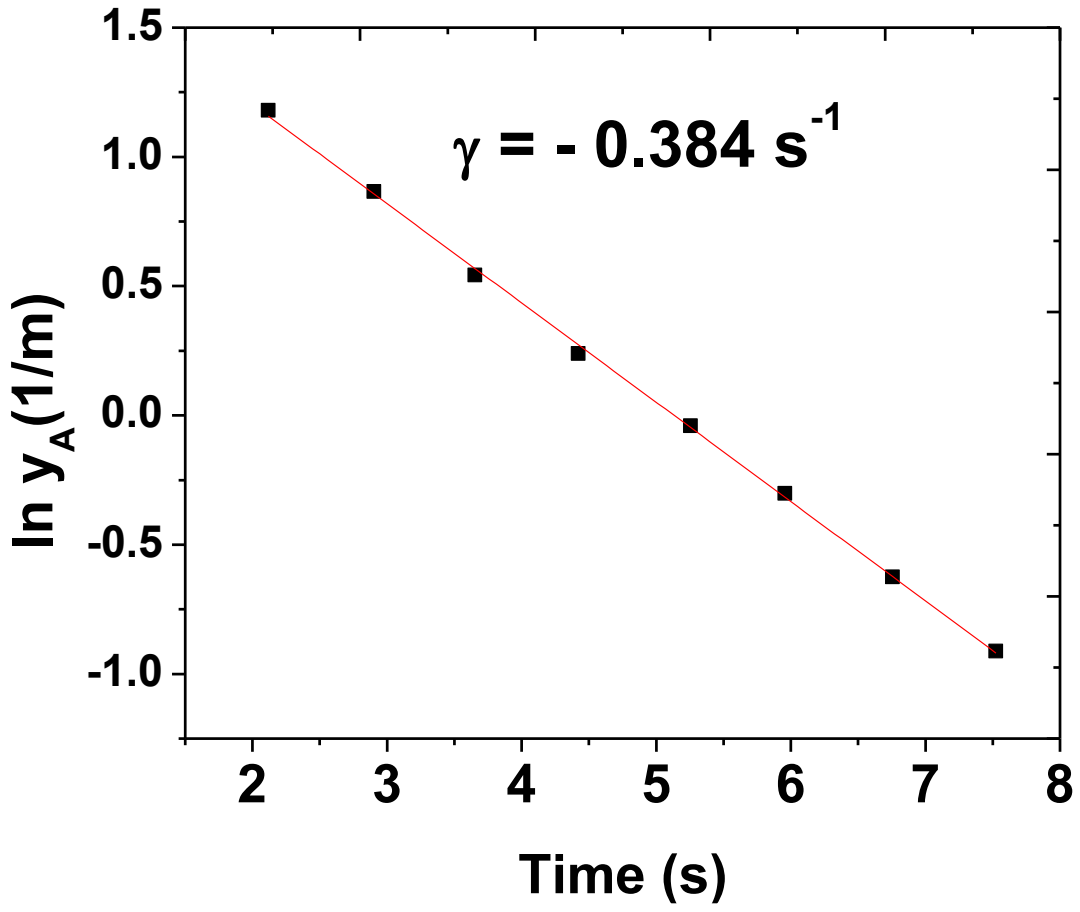

Fig 4: ln $y_A$ vs time graph for tracker application method

The Tracker application provides a graph showing the variation of amplitude with time. We have noted the coordinates of the peak points over time and plotted a graph of ln ($y_A$) versus time (see Fig 4). The slope of the straight-line graph is -0.384 $s^{-1}$, which gives the value of damping constant *b* is 14.803 gm.$s^{-1}$. Using this value of *b*, we have determined the coefficient of viscosity of the mustard oil as 0.625 poise.

## Conclusion

In conclusion, this study presents a straightforward method for measuring liquid viscosity using a smartphone's accelerometer. The coefficient of viscosity for mustard oil, as determined through damped oscillation analysis via the Phyphox and Tracker apps, shows close agreement with standard reference values, demonstrating the accuracy of the approach. This work underscores the practicality of using smartphone sensors for precise and cost-effective measurements, positioning this method as a valuable tool for both educational experiments and research applications.

## References

1. S. K. Pal, S. Sarkar, and P. Panchadhyayee, "Smartphone-based measurement of magnetic force and demonstration of Newton's third law of motion," *Phys. Teach.* **62,** 404-405 (2024).
2. S. K. Pal, S. Sarkar, and P. Panchadhyayee, "LiDAR based determination of spring constant using smartphones," *The Phys. Educat.* **6**, 2450001 (2024).
3. P. Vogt and J. Kuhn, "Analyzing simple pendulum phenomena with a smartphone acceleration sensor," *Phys. Teach.* **50**,439–440 (2012).
4. P. Vogt and J. Kuhn, "Analyzing free fall with a smartphone acceleration sensor," *Phys. Teach.* **50**, 182–183 (2012).
5. S. K. Pal, S. Sarkar, and P. Panchadhyayee, "Determination of the magnetic moment of a magnet by letting it fall through a conducting pipe," *Phys. Educ*. **59**, 015022 (2024).
6. S. K. Pal, S. Sarkar, and P. Panchadhyayee, "Determination of the Acceleration due to Gravity by Studying the Magnet's Motion Through a Conducting Pipe," *The Phys. Educat.* **6**, 2450008 (2024).
7. J. Kuhn and P. Vogt, "Analyzing spring pendulum phenomena with a smartphone acceleration sensor," *Phys. Teach*. **50**, 504 (2012).
8. J J Mendoza-Arenas, E L D Perico, and F Fajardo, "Motion of a damped oscillating sphere as a function of the medium viscosity," *Eur. J. Phys.* **31**, 129–141 (2010)
9. S. K. Pal, S. Sarkar, P. Panchadhyayee, and D. Syam, "Study of the effect of electromagnetic damping force on a magnet oscillating near a non-ferromagnetic conducting plate," *Eur. J. Phys.* **45**, 045203 (2024).
10. A. K. Azad, S. M. A. Uddin, and M. M. Alam, "A comprehensive study of di diesel engine performance with vegetable oil: an alternative bio fuel source of energy," *International Journal of Automotive and Mechanical Engineering.* **5**, 576-586 (2012)